\documentclass[prb,twocolumn,amsmath,amssymb,showpacs]{revtex4-1} %fjern 1 tal hvis gammel revtex4 version
\usepackage{graphics}
\usepackage{dcolumn}% Align table columns on decimal point
\usepackage{bm}% bold math
\usepackage{color}
\usepackage{epstopdf}
\usepackage{epsfig}

\begin{document}

\title{Glassy low-energy spin fluctuations and anisotropy gap in La$_{1.88}$Sr$_{0.12}$CuO$_4$}

\author{A. T. R\o mer,$^{1,2}$ J. Chang,$^{3,4,5}$ N. B. Christensen,$^{6}$ B. M. Andersen,$^1$ 
K. Lefmann,$^{1}$ L. M\"ahler,$^3$ J. Gavilano,$^3$ R. Gilardi,$^3$
Ch. Niedermayer,$^3$ H. M. R\o nnow,$^2$ A. Schneidewind,$^7$ P. Link,$^8$ M. Oda,$^{9}$ M. Ido,$^{9}$ N. Momono,$^9$ and J. Mesot$^{3,4,5}$}
\affiliation{$^1$Niels Bohr Institute, University of Copenhagen, DK-2100 Copenhagen, Denmark\\
$^2$Laboratory for Quantum Magnetism, \'Ecole Polytechnique F\'ed\'erale de Lausanne (EPFL), 1015 Lausanne, Switzerland\\
$^3$Laboratory of Neutron Scattering,  Paul Scherrer Institute, 5232-Villigen, Switzerland\\
$^4$Laboratory for Solid State Physics, ETH Z\"urich, CH-8093 Z\"urich, Switzerland\\
$^5$Institut de la materi\`ere complexe, \'Ecole Polytechnique F\'ed\'erale de Lausanne (EPFL), 1015 Lausanne, Switzerland\\
$^6$Department of Physics, Technical University of Denmark, DK-2800 Kgs. Lyngby, Denmark\\
$^7$Institut f\"ur Festk\"orperphysik (IFP), Technische Universit\"at 
Dresden, D-01062 Dresden, Germany\\
$^8$Forschungsneutronenquelle Heinz Maier-Leibnitz (FRM II), 
Technische Universit\"at M\"unchen, 85747 Garching, Germany\\
$^9$Department of Physics, Hokkaido University, Sapporo 060-0810, Japan}

\begin{abstract}
We present high-resolution triple-axis neutron scattering studies of the high-temperature superconductor La$_{1.88}$Sr$_{0.12}$CuO$_4$ ($T_c=27$ K). The temperature dependence of the low-energy incommensurate magnetic fluctuations reveals distinctly glassy features. The glassiness is confirmed by the difference between the ordering temperature $T_N\simeq T_c$ inferred from elastic neutron scattering and the freezing temperature $T_f \simeq 11$ K obtained from muon spin rotation studies. The magnetic field independence of the observed excitation spectrum as well as the observation of a partial suppression of magnetic spectral weight below $0.75$ meV for temperatures smaller than $T_f$, indicate that the stripe frozen state is capable of supporting a spin anisotropy gap, of a magnitude similar to that observed in the spin and charge stripe ordered ground state of La$_{1.875}$Ba$_{0.125}$CuO$_4$. The difference between $T_N$ and $T_f$ implies that the significant enhancement in a magnetic field of nominally elastic incommensurate scattering is caused by strictly in-elastic scattering -- at least in the temperature range between $T_f$ and $T_c$ -- which is not resolved in the present experiment. Combining the results obtained from our study of La$_{1.88}$Sr$_{0.12}$CuO$_4$ with a critical reappraisal of published neutron scattering work on samples with chemical composition close to $p=0.12$, where local probes indicate a sharp maximum in $T_f(p)$, we arrive at the view that the low-energy fluctuations are strongly dependent on composition in this regime, with anisotropy gaps dominating only sufficiently close to $p=0.12$ and superconducting spin gaps dominating elsewhere.
\end{abstract}


\maketitle

\section{Introduction}
In the presence of quenched disorder, competing order parameters in strongly correlated electron systems
are known to result in interesting physical phenomena such as phase separation, glassiness, and
dramatic responses to applied stimuli.~\cite{Dagotto05}
Doped transition metal oxides are perhaps the most studied examples of these general themes. Notably high-temperature
superconducting cuprates have attracted enormous interest since their discovery.~\cite{Bednorz86}  
Undoped cuprates are charge transfer insulators  that upon charge-carrier doping of the CuO$_2$ layers become superconducting. 
For carrier concentrations lower than optimal for superconductivity, {\it i.e.} in the underdoped 
regime, several competing or coexisting order parameters have been identified, such as circulating orbital 
currents,~\cite{Fauque06,Li08,Baledent10} incommensurate spin and charge stripe ordering~\cite{Tranquada95, Fujita04} and, recently, 
charge density wave order.~\cite{Ghiringhelli12,Chang_natphys_12}
Stripe order in cuprates was originally discovered when the hole-doping level $p$ was 
tuned to $p=x=1/8$ in La$_{1.6-x}$Nd$_{0.4}$Sr$_{x}$CuO$_{4}$ (Nd-LSCO).~\cite{Tranquada95}
Subsequently, stripe order has been observed in La$_{2-x}$Ba$_{x}$CuO$_{4}$ (LBCO)~\cite{Fujita04,Hucker11}
and La$_{1.8-x}$Eu$_{0.2}$Sr$_{x}$CuO$_{4}$ (Eu-LSCO)~\cite{Klauss2000,Fink2009} -- also at $p=1/8$. 
In all three cases, competition between incommensurate spin-charge order and superconductivity 
causes a dramatic drop of $T_c$, which reaches very low values at $x=1/8$, where the stripe ordering 
tendencies are most pronounced.
Even in the archetypal superconductor YBa$_{2}$Cu$_{3}$O$_{y}$ (YBCO) that has 
an optimal $T_c = 90$ K, a small suppression of the superconducting transition 
temperature has been found near the 1/8 doping.~\cite{Liang06} In the case 
of YBCO the exact nature of the competing order parameter is still being explored, 
with the most recent evidence from NMR,~\cite{Wu_Nat_11} transport,~\cite{LaLiberte_NatCom_11} and x-ray scattering techniques~\cite{Ghiringhelli12,Chang_natphys_12} pointing to charge-density wave order.
La$_{2-x}$Sr$_{x}$CuO$_{4}$ (LSCO) falls in between YBCO and the stripe compounds Nd-LSCO and LBCO.
Like YBCO, it displays only a small suppression of $T_c$ in the vicinity of $p=x=1/8$, \cite{Takagi89} 
but in this regime, incommensurate magnetism similar to that found for Nd-LSCO and LBCO coexists with 
superconductivity in the ground state.~\cite{Suzuki98}
To date, no evidence for incommensurate bulk charge order in LSCO has emerged.~\cite{Wu_NatCom12} 
This difference is believed to be an effect of a favorable potential for charge-stripe pinning in the specific low-temperature 
tetragonal (LTT) structure of Nd-LSCO and LBCO. Within this picture, the low-temperature orthorhombic (LTO) structure of LSCO is not suitable for charge stripe order, but does allow incommensurate magnetism near $x=1/8$.~\cite{Suzuki98,Chang08} The onset temperature of the incommensurate magnetism in LSCO depends on the experimental technique used to probe it. This implies that the electronic spins are gradually freezing rather than undergoing a regular thermodynamic phase transition. For LSCO the freezing temperature, $T_f$, derived from local probes such as NMR, NQR and muon spin rotation ($\mu$SR) has a narrow peak centred around $x_{\textit max} \sim 0.12$.~\cite{MHJulien_physicaB}

It appears reasonable to conjecture that the details of the magnetic excitation spectrum may be highly sensitive to doping near $x_{max}$. The available experimental evidence is, however, limited and a consistent interpretation is lacking.
A recent doping-dependent study of the low-energy dynamics in LSCO by M. Kofu {\it et al. }~\cite{Kofu09} reported a correlation between the presence of incommensurate elastic magnetic scattering and gapless spin excitations near  $x_{max}$. These data 
were interpreted in terms of two components: a spin-gapped response similar to what is observed at optimal doping,~\cite{Mason92} {\it i.e.} a gap caused by superconductivity, and a second component related to spin/charge stripe-ordered or stripe-frozen domains.
In contrast to this view, J. Chang {\it et al.}~\cite{Chang07} proposed that for $x=0.105<x_{\textit max}$
magnetic order renormalizes the value of the superconductivity-related
spin gap. The two-component view is also in contrast to what is observed at $x=0.145 > x_{max}$ where the spin gap is found to close at the quantum critical point for the incommensurate spin order.~\cite{Chang09,Christensen11}

Here we present results of experiments designed to improve our understanding of the low-energy dynamic magnetic fluctuations in LSCO and move towards a consistent description of its doping-dependence near $x_{max}$. We have studied the temperature and magnetic field-dependence of the spin-dynamics of LSCO at $x=0.12 \simeq x_{max}$. 
In contrast to the clear effects of magnetic field on low-energy excitations reported both for the underdoped regime for $x < x_{max}$~\cite{Chang07} and for the optimally doped regime for $x > x_{max}$,~\cite{Lake01,Tranquada04,Gilardi04,Chang09} we observe no field-effect for $x\simeq x_{\textit max}$. The temperature-dependence of the spin fluctuations reveal glassy dynamics and, at the lowest energy transfers, a partial suppression of magnetic spectral weight below $0.75$ meV. The latter observation in combination with the absence of a magnetic field effect on the low-energy excitations, suggests that LSCO near $x_{\textit max}$ can support a spin anisotropy gap despite the glassy nature of the ordering and despite superconductivity. We discuss the implications of our results and their relation to previously published work on La$_{2-x}$Sr$_{x}$CuO$_{4}$, La$_{2-x}$Ba$_{x}$CuO$_{4}$ and 
spin-charge ordered nickelates La$_{2-x}$Sr$_{x}$NiO$_{4}$ in section \ref{section_dis}. Here we arrive at a validation of the conjecture that the low-energy excitations are strongly dependent on hole-content near the maximum in $T_f$. First, however, we describe the experimental methods in section \ref{section_exp} and present our experimental data in section \ref{section_res}. 

\begin{figure}
\includegraphics[clip=true,width=0.98\columnwidth]{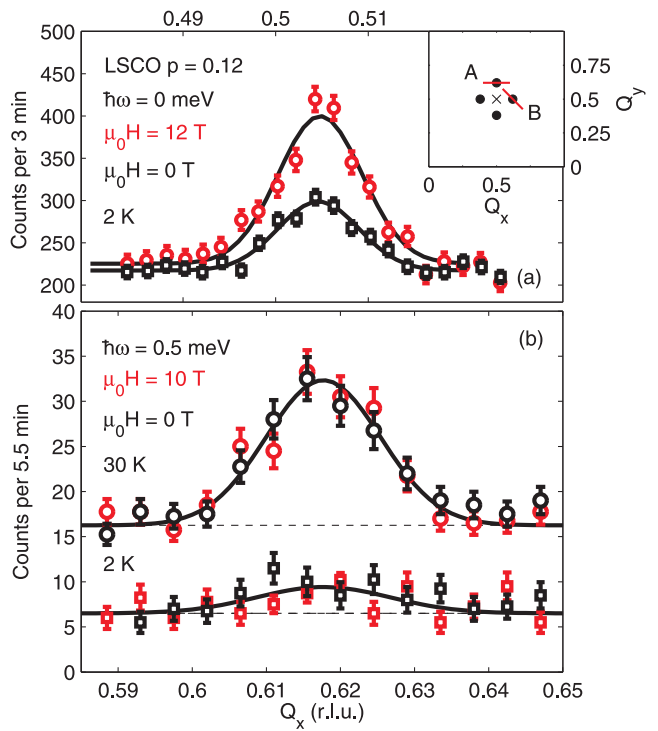}
\caption{(a) Elastic scattering intensity at the incommensurate position $Q_{\rm IC}$ as indicated 
in the inset where circles mark the locii of incommensurate magnetic order and low-energy fluctuations. Black and red data points were taken at 0 T and 12 T field, respectively. Solid lines are Gaussian fits to data. These data are reproduced from Ref.~\onlinecite{Chang08}. The inset shows the two different scan directions for the elastic scattering data (A) and the inelastic scattering data (B).
(b) Background-subtracted intensity for a constant energy scan with $\hbar \omega=0.5 $ meV. The data has been shifted upwards by a constant offset for clarity.
Solid lines are Gaussian fit to data. 
The application of a magnetic field of 10 T has no effect on 
the low-energy spin excitations ($\hbar\omega = 0.5$ meV) neither 
for $T\ll T_{\rm c}$ nor in the normal state $T=30$~K $> T_{\rm c}$. Notice that 
for visibility, the 30 K data have an arbitrary offset. The nominally elastic peak and the inelastic peaks are all resolution limited; $\xi_{\rm elastic} \geq 110$ \AA ~and $\xi_{0.5 {\rm ~meV}} \geq 70$ \AA.}
\label{fig:Fig1}
\end{figure}

\section{Experimental method}
\label{section_exp}
The La$_{1.88}$Sr$_{0.12}$CuO$_4$ ($T_{\rm c} \approx 27$ K) sample studied consisted of two single crystals which were cut from the same rod grown
by the travelling solvent floating method.~\cite{Nakano98} In earlier work,~\cite{Chang08} the Sr content $x=0.120 \pm0.005$ (and hence the hole concentration $p=x$) was determined from the structural transition temperature separating the high-temperature tetragonal (HTT) from the low-temperature orthorhombic (LTO) phase.
Muon spin rotation studies on one of the two single crystals revealed electronic moments that are static on the muon time scale below a freezing temperature $T_f \simeq 11$ K.~\cite{Chang08,Larsen2012}
High-resolution inelastic neutron scattering experiments were carried out on the PANDA cold neutron triple axis spectrometer at the FRM-II research neutron source in Munich, Germany. 
In a first experiment, the two rods were co-aligned to within less than one degree and the sample was inserted in a 15~T vertical field cryomagnet and the instrument configured with vertically focusing pyrolytic graphite (PG) monochromator and collimation sequence open-60{$^\prime$}-open-open
from source to detector. In a second, zero-field, experiment one rod was used in a setup with a double-focusing monochromator and no collimation.
In both experiments, a cooled Be-filter was placed in front of the double-focusing PG analyzer to minimize contamination from higher-order neutrons. The sample was oriented with the crystallographic c axis vertical, allowing access to wave vectors of the form $Q=(H,K,0)$. In labelling reciprocal space, it is convenient to use notation corresponding to the high-temperature tetragonal crystal structure ($a \simeq b = 3.78$ \AA, $c \simeq 13.18$ \AA). In this notation, the propagation vector of the undoped antiferromagnetic parent compound La$_2$CuO$_4$~\cite{Vaknin87} is $(1/2,1/2,0)$ while stripe magnetic ordering~\cite{Tranquada95} is manifested in a quartet of peaks, incommensurate with the lattice at $Q_{\rm IC}=(1/2 \pm \delta,1/2,L)$ and $(1/2 ,1/2 \pm \delta,L)$ (See the inset in Fig.~\ref{fig:Fig1}(a)). For our sample, $\delta = 0.125(3)$ as reported earlier.~\cite{Chang08} 
The intensity recorded by the neutron detector is the convolution of the instrumental resolution function with the spin-spin correlation function $S({\bf Q},\omega)$ which in turn is related to the imaginary part of the generalized magnetic susceptibility $\chi^{\prime\prime}({\bf Q},\omega)$ via the fluctuation dissipation theorem
\begin{equation}\label{eq1}
S({\bf Q},\omega) = \chi^{\prime\prime}({\bf Q},\omega,T)n_B(\omega,T).
\end{equation}
where $n_B(\omega,T)=(1-e^{-\hbar\omega/k_{\rm B}T})^{-1}$ is the Bose occupation factor.
In addition to the magnetic scattering $S({\bf Q},\omega)$, the raw
experimental data also contain contributions
which do not arise from electronic magnetism, but are due to incoherent
scattering
from atomic nuclei.
To obtain $\chi^{\prime \prime}(Q,\omega,T)$ it is important to
cleanly separate these contributions.
For the strongly peaked response observed at low energies in LSCO, an
effective background subtraction procedure
is to estimate the non-magnetic contributions from the scattering observed
at wave vectors sufficiently far away
from the magnetic peaks.
We studied the magnetic fluctuations over the temperature range 2-80 K and for energy transfers in the range 0.3-7 meV. Most of the results we report were obtained with fixed final neutron energy $E_{\rm f}=5.0$~meV. For measurements of spin excitations at energy transfers, $\hbar\omega=$0.3-0.5~meV
we chose a lower final energy $E_{\rm f}=4.1$~meV to avoid contamination from strictly elastic scattering through the finite energy resolution. In this case the energy resolution was $0.13$ meV FWHM as compared to $0.18$ meV at $E_f=5.0$~meV.

\begin{figure}
\begin{center}
\includegraphics[clip=true,width=0.98\columnwidth]{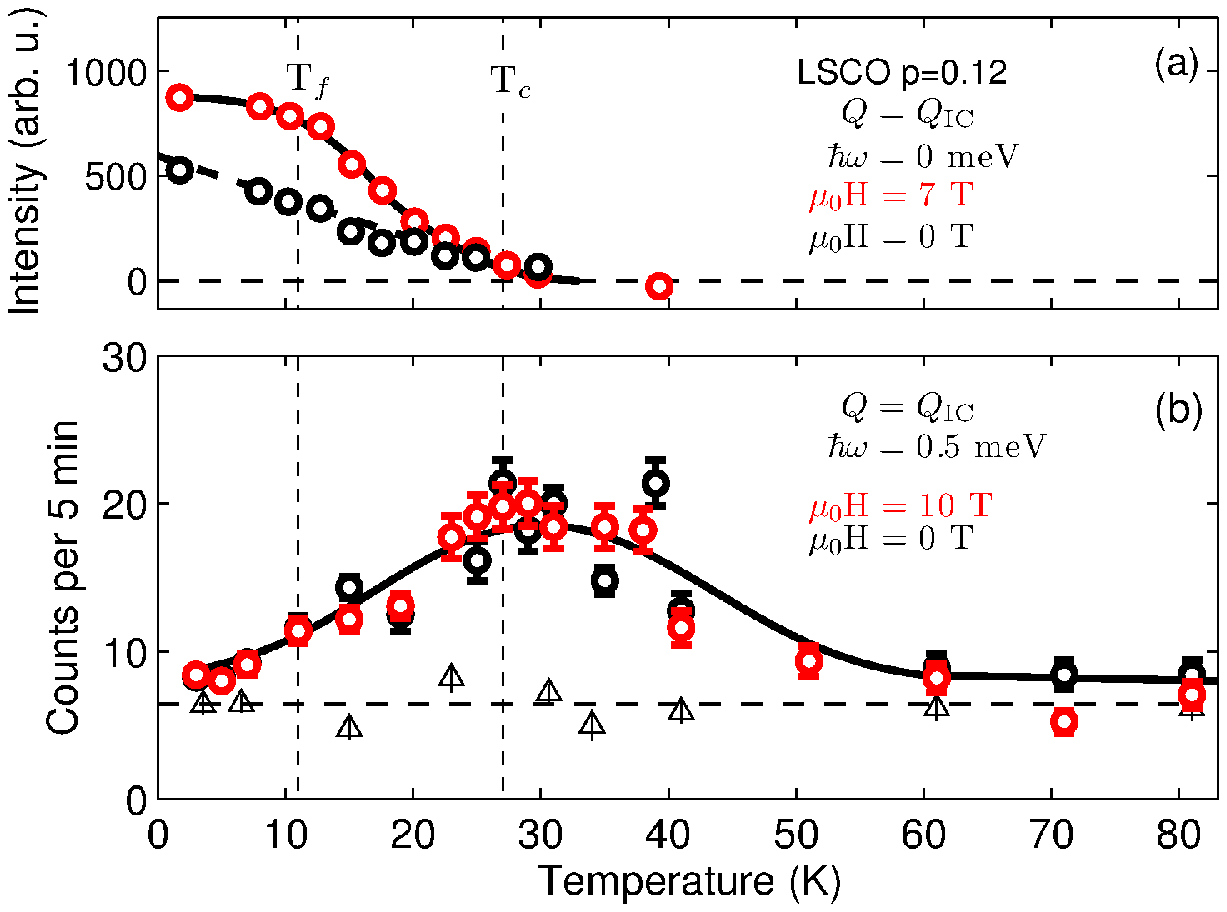}
\includegraphics[clip=true,width=0.98\columnwidth]{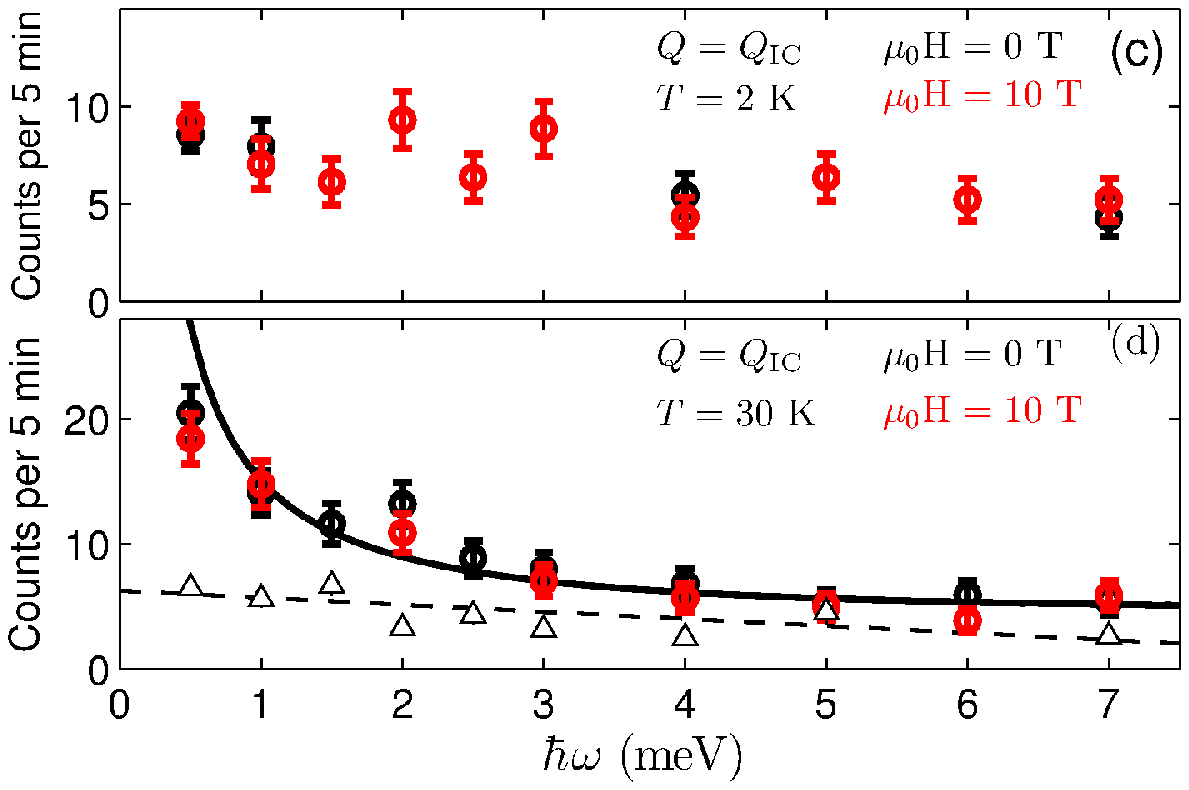}
\end{center}
\caption{(a) Temperature dependence of the background-subtracted elastic response at $Q_{\rm IC}$ for $H=0$~T (black) and 7~T (red). The vertical dashed lines indicate the superconducting transition temperature, $T_c$=27 K and the freezing temperature for magnetic ordering, $T_f \sim 11$ K, obtained from muon spin rotation.~\cite{Larsen2012} (b) Temperature dependence of the inelastic response (0.5~meV) at $Q_{\rm IC}$ in zero field (black) and 10~T (red). Open triangles are background data.
(c) and (d) Inelastic response at $Q_{\rm IC}$ as a function of energy in the superconducting state at $T=2$~K and normal phase  at $T=30$~K both with (red points) and without (black points) an applied field of 10~T. Open triangles are background data. The dashed line is a fit to a linear function. The solid black line in (d) is a fit to the Bose occupation factor as described in the text.}
\label{fig:Fig2}
\end{figure}

\begin{figure}
\includegraphics[clip=true,width=0.95\columnwidth]{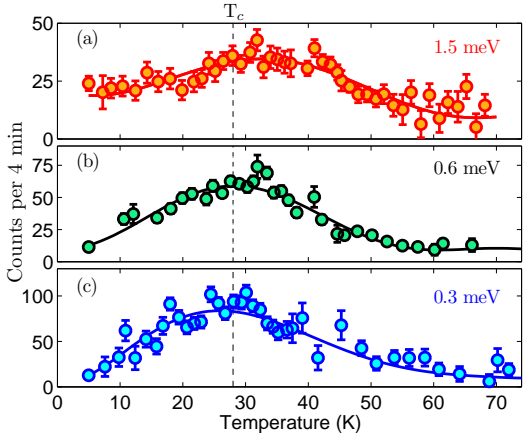}
\caption{Background-subtracted inelastic response at $Q = Q_{\rm IC}$ in zero field 
and plotted as intensity versus temperature for excitation energies: (a) 1.5 meV, (b) 0.6 meV, 
and (c) 0.3~meV. 
Vertical dashed line indicates the $T_{\rm c}$ of this compound and solid lines 
are guides to the eye. Since these measurements were obtained without a magnet there is a significant increase in intensity compared to the data shown in Figs.~\ref{fig:Fig1} and ~\ref{fig:Fig2}.}
\label{fig:Fig3}
\end{figure}

\section{Results}
\label{section_res}
The temperature and magnetic field dependence of static magnetism in LSCO $p\sim0.12$, as well as its momentum space characteristics has been previously studied in great detail.~\cite{Suzuki98,Kimura00,Katano00,Chang08,Kofu09}
In Fig.~\ref{fig:Fig1}(a) we show a constant energy scan obtained with the spectrometer set to energy transfer $\hbar\omega=0$ meV. Momentum-resolution limited peaks are observed close to $(1/2,1/2 + \delta, 0)$.
The slight offset is consistent with the observation of Kimura {\it et al.}~\cite{Kimura00} that the incommensurate, nominally elastic peaks do not lie along the high-symmetry directions of the underlying CuO$_2$ lattice, but are slightly displaced.
In our sample, the magnetic intensity increases significantly when a magnetic field is applied along the crystallographic c-axis, see Fig.~\ref{fig:Fig1}(a) and Ref.~\onlinecite{Chang08}. The onset temperature for the magnetic order is essentially field-independent, as can be seen in Fig.~\ref{fig:Fig2}(a). 

\begin{figure}
\includegraphics[clip=true,width=0.98\columnwidth]{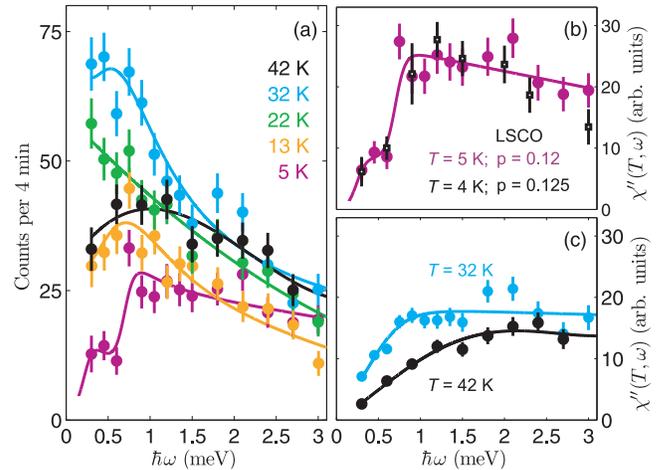}
\caption{(a) Background-subtracted inelastic response as a function of excitation energy at different temperatures $T$. (b) Dynamical susceptibility $\chi^{\prime\prime}(T,\omega)$ at  $T=5$ K obtained for our $p=0.12$ sample and compared to low-temperature data reproduced from M. Kofu {\it et al}.~\cite{Kofu09} for their $p=0.125$ sample at $T=4$ K  (c) Dynamical susceptibility $\chi^{\prime\prime}(T,\omega)$ for fixed $T=$ 32 K and 42 K. All lines are guides to the eye.}
\label{fig:Fig4}
\end{figure}

Figure ~\ref{fig:Fig1}(b) displays the inelastic response at $Q_{\rm IC}$ with energy transfer $\hbar\omega=0.5$~meV,  
probed at base temperature and just above $T_c$. Data taken in zero field and at $H=10$~T are shown.
In strong contrast to the elastic response shown in Fig.~\ref{fig:Fig1}(a), no detectable field effect was observed at any temperature, 
as seen in Fig.~\ref{fig:Fig2}(b).
The lack of field-effect persists throughout the energy range $0.3$-$7.0$ meV at both $T=2$ K and $T=30$ K  as shown in Figs.~\ref{fig:Fig2}(c) and~\ref{fig:Fig2}(d), respectively. 
The peak position was determined by full $Q$-scans as in Fig.~\ref{fig:Fig1}(b) and was observed to be independent of temperature and energy transfer within the temperature and energy range of this experiment. From the $Q$-scans we found that the inelastic correlation length is resolution limited by $\xi(T,\omega)\approx 70$~\AA\ for $\hbar\omega<3$ meV and $T<50$~K. 
Further data were therefore taken by three-point scans; counting at the peak position and two background positions on each side of the peak.
The background estimates are subtracted from the peak intensities in Figs. \ref{fig:Fig3}(a-c) and \ref{fig:Fig4}(a). 
The solid line in Fig.~\ref{fig:Fig2}(d) is the Bose occupation factor $n_B(\omega,T)$ scaled to the data. This lead us to the conclusion that the energy dependence of $S({\bf Q},\omega)$ at $30$ K is dominantly given by the Bose occupation factor for energies in the range $1$-$7$ meV, and hence that $\chi^{\prime \prime}$ is roughly frequency independent. 

To elucidate the details of the temperature dependence, we show in Fig.~\ref{fig:Fig3} the temperature dependence of the inelastic response for three different energies from 0.3~meV to 1.5~meV. 
As in Fig.~\ref{fig:Fig2}(b) we observe a broad maximum around $T_{\rm c}$. The position of the maximum shifts down in temperature as $\hbar\omega$ is decreased, and approaches $T_c$ in the limit  $\hbar \omega \rightarrow 0$ meV. 
For all energy transfers probed, the intensity decreases as the sample is cooled from $T_c$ to base temperature. 
This tendency is much less pronounced for $1.5$ meV than for $0.3$ meV. At the former, the intensity remains finite in the low-temperature limit, whereas it approaches zero for the latter.

To investigate how the suppression of intensity at very low energies ($\hbar \omega <1$ meV) manifests itself in the observed spectra, we show in Fig.~\ref{fig:Fig4} the energy dependence of the incommensurate signal for several temperatures above and below $T_{\rm c}$. Figure~\ref{fig:Fig4}(a) shows how the intensity is drastically reduced for energy tranfers lower than $0.75$~meV at $T=5$~K and, to a lesser extent, for $T=13$~K. This shows that $\chi^{\prime\prime}(T,\omega)$ becomes frequency-dependent for temperatures lower than and comparable to the freezing temperature $T_f\simeq 11$ K deduced from $\mu$SR. The guide to the eye for the 5 K data suggests an interpretation in terms of two energy gaps. We return to this point in the discussion.

To illustrate the effects of the Bose occupation factor, see Eq.~(\ref{eq1}), we plot the corresponding dynamic 
susceptibilities in Figs.~\ref{fig:Fig4}(b) and~\ref{fig:Fig4}(c) for three temperatures, $5$ K, and $32$~K and $42$~K. By contrast to the smooth energy dependence of the susceptibility at high temperatures, the $5$~K data display an abrupt reduction by a factor of roughly four below $\hbar \omega \sim 0.75$~meV, see Fig.~\ref{fig:Fig4}(b). This observation is consistent with measurements at a similar doping level, $p=0.125$, by M. Kofu {\it et al}.~\cite{Kofu09}. We stress that the intensity suppression is only partial even at the lowest frequency, as seen in Figs.~\ref{fig:Fig3}(c) and ~\ref{fig:Fig4}(b), but signatures of gapping of the spectrum are clear from Fig.~\ref{fig:Fig4}, Fig.~\ref{fig:Fig3}(b,c), and Fig.~\ref{fig:Fig2}(b). Moreover, from Fig.~\ref{fig:Fig2}(b), it is apparent that these gap signatures do not exhibit any magnetic field dependence. 
 
\section{Discussion}
\label{section_dis}
In this section we first discuss the temperature dependence of the inelastic data and indications of glassy dynamics. Thereafter we turn to the energy dependence and the intensity suppression observed at low energy transfers. Finally, we discuss the field dependence of the nominally elastic data and the field independence of the inelastic data.

\subsection{Freezing and glassiness} 
The peaked response of the low-energy fluctuations shown in Figs.~\ref{fig:Fig2}(b) and~\ref{fig:Fig3} is a common feature observed throughout the doping range in LSCO.~\cite{Chang07,Yamada95,Lake01} We observe a peak in the $S({\bf Q},\omega)$ at temperatures close to $T_c$ which shifts towards higher temperatures as the energy transfer, $\hbar\omega$, increases. This shift, consistent with glassy behaviour, was also observed in very underdoped, non-superconducting LSCO $p=0.04$ (Ref. \onlinecite{Keimer92}). In that case the integrated spin intensity peaked at a temperature which increased for increasing energy; the peak temperature was given by $T \sim 2 \hbar \omega$. In our case we probe only low energy transfers, but the tendency is the same, {\it i.e.} the threshold temperature above which the intensity drops off increases with frequency, see Fig.~\ref{fig:Fig3}. 
We stress that since the
observed peak widths are roughly constant for the
range of temperatures and energy transfers probed in our experiments, the
peak amplitude at $Q_{\rm IC}$ is to a good approximation
proportional to the integrated intensity discussed in Ref.~\onlinecite{Keimer92}. We
therefore conclude that our La$_{1.88}$Sr$_{0.12}$CuO$_4$
 crystal displays
low-energy dynamics that are similar to what was found in the glassy
ground state of very underdoped LSCO $p=0.04$.~\cite{Keimer92,Chou95,Wakimoto00}

Turning to magnetic order, we observe a nominally elastic signal for $T<T_c$, see Fig.~\ref{fig:Fig2}. For an energy resolution of $\Delta E\sim 0.18$ meV the time resolution $t\sim \frac{\hbar}{\Delta E} $ is of the order tens of picoseconds. Fluctuations with a characteristic time scale larger than picoseconds will therefore appear as static. For our crystal the freezing temperature obtained by zero field $\mu$SR, for which the time resolution is of the order of microseconds, is $T_f \sim 11$ K.~\cite{Larsen2012} 
This implies that the magnetic ordering temperature as observed by neutron scattering is comparable to $T_c$ only by a coincidence. Further, it implies that the nominally elastic neutron scattering signal observed in the temperature range between $T_f$ and $T_c$ is actually caused by strictly inelastic, low-energy magnetic fluctuations picked up by the experimental resolution function of the spectrometer, {\it i.e.} fluctuations with characteristic energy scale $0.18$ meV or lower. Note also that $T_f$ obtained by $\mu$SR only sets an upper limit of the actual freezing temperature: An experimental technique probing the spin dynamics on a longer characteristic timescale than microseconds could give an even lower value.

\subsection{Anisotropy gap}
Figure~\ref{fig:Fig4} shows that for temperatures lower than $T_f$, we observe a partial suppression of low-energy fluctuations. There is an intensity drop at energy transfers lower than $\hbar \omega \sim 0.75$ meV. Below this scale, we do not see a spectral region of zero intensity, which means that the gap is not fully developed. We now turn to discuss the possible origin of these observations within a simple spin wave formalism.
The energy gaps in the parent compound La$_2$CuO$_4$ are due to exchange anisotropy and the Dzyaloshinskii-Moriya interaction. These gaps were reported by C. J. Peters {\it et al.}~\cite{Peters88} to be 1.0~meV and 2.5~meV for the in-plane and out-of-plane gaps, respectively. Within the standard Heisenberg spin-only approach, an anisotropy gap is expected to scale with the ordered magnetic moment of the Cu atoms. This is a generic result also expected to hold for stripe ordered systems for which the static moments are known to be strongly diminished compared to La$_2$CuO$_4$.~\cite{Wakimoto01}
The ordered moment in our sample was previously determined to be an order of magnitude smaller than that of La$_2$CuO$_4$.~\cite{Chang08} Therefore, ignoring any additional effects due to quenched disorder produced by the replacement of La by Sr, we can expect anisotropy gaps in La$_{1.88}$Sr$_{0.12}$CuO$_4$ to be roughly an order of magnitude smaller than in La$_2$CuO$_4$. This would make the gaps comparable to the energy resolution of our experiments.
An interpretation of our data which is consistent with these qualitative arguments is that we observe the out-of-plane energy gap at $\hbar \omega \sim 0.75$~meV, while our experiment does not resolve the smaller in-plane gap. 
This explains why we observe only a partial suppression of the scattering signal rather than a fully developed energy gap.
Note, however, that experiments probing spin fluctuations at a single incommensurate wave vector only do not allow us to directly identify 
the larger gap as due to out-of-plane anisotropy rather than in-plane anisotropy. 

Evidence for a residual small spin anisotropy gap was also discussed for a $p=0.04$ sample in the non-superconducting spin-glass regime of LSCO.~\cite{Keimer92} Turning to stripe-ordered La$_{1.875}$Ba$_{0.125}$CuO$_4$, a low energy intensity suppression of the same magnitude below $0.7$~meV, was recently observed~\cite{Tranquada08,Wen08} and similarly ascribed to spin anisotropy of the spin-ordered state. In that case, too, a magnetic field effect of the low-energy excitations was absent. Hence, we find striking similarities for the low-energy excitations between our sample and La$_{1.875}$Ba$_{0.125}$CuO$_4$, indicating that both samples have similar magnetic regions and anisotropy gaps. 

The nickelates, La$_{2-x}$Sr$_{x}$NiO$_{4}$, also show a stripe ordered phase upon doping with regions of antiferromagnetically ordered spins separated by parallel lines of holes. Although nickelates do not become superconducting upon doping and also do not display the hourglass dispersion common to 
La-based cuprates~\cite{Tranquada04,Christensen04} and La$_{2-x}$Sr$_{x}$CoO$_{4}$,~\cite{Boothroyd2011}, the existence of 
stripe order in both justify a qualitative comparison.
Studies of the low-energy magnetic dynamics in nickelates over a broad range of Sr content have shown evidence of an out-of-plane anisotropy gap which decreases with increasing doping.~\cite{Boothroyd2003,Boothroyd2004} The decrease is about a factor of two comparing the parent compound La$_{2}$NiO$_{4}$~\cite{Nakajima93} to doping $x = 0.275-0.37$ (See Ref. ~\onlinecite{Boothroyd2004}) and roughly a factor of three for doping values $x=0.4-0.45$.~\cite{PaulComm} This trend is similar to the behaviour we have identified in La$_{1.88}$Sr$_{0.12}$CuO$_4$.
Summarizing the above, it appears that the low-energy spin dynamics in underdoped La-based cuprates and nickelates can display small residual anisotropy gaps, irrespective of whether they are stripe-ordered as La$_{1.875}$Ba$_{0.125}$CuO$_4$~\cite{Tranquada08,Wen08} and La$_{2-x}$Sr$_{x}$NiO$_{4}$ or undergoing glassy freezing as in La$_{2-x}$Sr$_{x}$CuO$_4$. 

M. Kofu {\it et al.}\cite{Kofu09} recently reported a study of the low-energy excitations in LSCO at doping values $p=0.125-0.14$. The results were discussed in terms of a two-component scenario with a real space separation of two phases for $p \leq 0.13$. In the latter regime, spin fluctuations of short correlation length were proposed to exist at energies above an energy scale $E_g$ comparable to the spin gap observed at optimal doping, and to coexist with low-energy fluctuations which have significantly longer correlations length. Our data do not allow us to confirm or dismiss a change in correlation length with energy transfer. We note, on the other hand, that a partial suppression of low-energy spectral weight below an energy scale $\sim 1.0$ meV was detected by M. Kofu {\it et al.} in their $p=0.125$ sample (See Fig ~\ref{fig:Fig4}(b)). Moreover, no suppression was seen for $p=0.13$,~\cite{Kofu09} which resides at the edge of magnetic order ~\cite{MHJulien_physicaB} and should therefore have 
a much smaller ordered moment and hence much smaller anisotropy gaps.  
Both observations are
consistent with our interpretation of the low-energy intensity suppression
as originating from an anisotropy gap.

\subsection{Magnetic field effect}
An intriguing difference becomes apparent when comparing the magnetic field effect of the nominally elastic (Figs.~\ref{fig:Fig1}a, \ref{fig:Fig2}a)
 and inelastic (Figs.~\ref{fig:Fig1}b,~\ref{fig:Fig2}b-d) signals. Due to the finite energy resolution of a neutron experiment we conclude from a comparison of Fig.~\ref{fig:Fig2}(a) with $\mu$SR results that there is a significant field effect in the very low-energy fluctuations ($\hbar \omega<0.18$ meV) for temperatures below $T \sim 25$ K. This effect stands in contrast to the field-independent magnetic response in the energy range $0.5-7$ meV. The absence of a field effect in this range is distinctly different from the behaviour observed for compositions with slightly smaller ~\cite{Chang07} as well as higher hole-doping. ~\cite{Lake01,Tranquada04,Gilardi04,Chang09} These samples exhibit a magnetic-field enhancement of the spectral weight at low energies for temperatures below $T_{\rm c}$. 

In a spin-only Heisenberg approach where we account for an easy axis and Dzyaloshinskii-Moriya anisotropies, we can estimate the effect of an applied magnetic field $H$ on the anisotropy gap at the antiferromagnetic ordering vector.  Due to the Dzyaloshinskii-Moriya interaction the spins tilt slightly out of the CuO$_2$ planes. In the mathematical expression for the energy gap, the applied magnetic field $H$ enters in a term which multiplies the small tilting angle. Therefore the energy gaps are only weakly dependent on $H$ and we estimate an energy change due to an applied field of $\mu_0H=10$ T of the same order of magnitude as our energy resolution. In conclusion, at the lowest temperatures, we do not necessarily expect to detect any significant effect of a magnetic field on the anisotropy gaps. This is consistent with our observations in La$_{1.88}$Sr$_{0.12}$CuO$_{4}$. 

Returning to the lowest energy ($\hbar\omega < 0.18$ meV) spin excitations in La$_{1.88}$Sr$_{0.12}$CuO$_{4}$ and their enhancement in a magnetic field, 
further experiments with improved energy resolution will be required to determine if they are of an origin distinct from the spin-wave like excitations observed at higher energies, or whether -- as the above arguments suggest -- they are related to the anisotropy gap not resolved by the present experiment.

The observation of a magnetic field-effect for $p=0.105$ and its interpretation as a renormalized superconducting spin-gap~\cite{Chang07}
may now be rationalized by the existence of a sharp peak around $x_{max}=0.12$ in the freezing temperature $T_f$.~\cite{MHJulien_physicaB} 
The peak in the freezing temperature is associated with an increased competition between static magnetic order and superconductivity and a decrease of $T_c$.~\cite{Takagi89} We have argued that for $p=0.12$, the spin-frozen low-temperature state permits an anisotropy gap akin to that observed in the parent insulator La$_2$CuO$_4$. Conversely, when the magnetic ordering/freezing tendencies are weakened by moving to away from $x_{max}$, the low-energy dynamics can be expected to become dominated by the physics of the  superconductor with its superconducting spin gap opening at $T_c$, rather than that of the insulator with its anisotropy gaps. This physical picture goes a long way towards reconciling the disagreements of interpretations between J. Chang {\it et al.}~\cite{Chang07} and M. Kofu {\it et al.}~\cite{Kofu09} 

\section{Conclusions}
We have studied the detailed temperature and energy dependence of low-energy magnetic fluctuations in La$_{1.88}$Sr$_{0.12}$CuO$_4$. 
A discrepancy between the magnetic ordering temperature derived by neutrons and muons shows that the spins undergo freezing rather than a regular phase transition. We find additional evidence for freezing in the temperature dependence of the low-energy fluctuations, which resembles the behavior observed in LSCO at much lower doping, in the non-superconducting, so-called spin-glass regime, and therefore conclude that even superconducting La$_{1.88}$Sr$_{0.12}$CuO$_4$ exhibits spin glass behavior.

Below the spin-freezing temperature $T_f \simeq 11$ K of our sample, we have detected an incomplete suppression of magnetic spectral weight at energies larger than our energy resolution, $\Delta E$, but smaller than $~0.75$ meV. We ascribe this effect to the development of a spin anisotropy gap in a spin frozen setting. Applying insights from spin wave theory, this interpretation is supported by the insensitivity, to within our experimental resolution, of the low-energy intensity suppression to an applied magnetic field of 10~T. It is notable, that the low-energy excitations in our La$_{1.88}$Sr$_{0.12}$CuO$_4$ sample are remarkably similar to those of La$_{1.875}$Ba$_{0.125}$CuO$_4$ which displays long-range spin and charge
stripe order. 

In contrast to the insensitivity of the anisotropy gap to applied magnetic field, a significant enhancement of nominally elastic, incommensurate magnetic signal was observed at temperatures lower than $T_{\rm c}$. Given that muon spin rotation yields a freezing temperature $T_f$ much smaller than $T_c$, the implication is that there must be a magnetic field effect on remnant spin excitations at energies inside our resolution window $\Delta E \simeq 0.18$ meV. 
 
Our experimental data and a comparison with published data on LBCO and LSCO have illuminated that the sharp maximum in the spin freezing temperature $T_f$ near $p=0.12$ is reflected equally dramatically in the lowest energy magnetic excitations, as the superconducting gap observed at optimal doping is replaced by a spin anisotropy gap sufficiently close to the maximum in $T_f$. 

\section*{Acknowledgements}
We acknowledge Paul G. Freeman for illuminating discussions on similarities between cuprates and nickelates.
This work was supported by the Swiss NSF (through NCCR, MaNEP, and grant no. 
200020-105151, PBEZP2-122855, PZ00P2-142434), by the Ministry of Education and Science of Japan. 
and by the Danish Council for Independent Research in Natural Sciences (FNU) through DANSCATT and the grant Magnetism in Superconductors. 
The present experiments were performed at the FRM-II research reactor 
and were supported  by the European Commission under 
the 7th Framework program through the "Research Infrastructures" action 
of the "Capacities" program, Contract no. CP-CSA INFRA-2008-1.1.1 Number 
226507-NMI3,

\end{document}